\documentclass[prl,aps,twocolumn,nofootinbib,preprintnumbers]{revtex4}
\usepackage{graphicx}
\usepackage{epstopdf}
\usepackage{epsfig,amssymb,amsmath,color}

\def\bea{\begin{eqnarray}}
\def\eea{\end{eqnarray}}

\begin{document}
\draft
\tighten
\preprint{PNUTP-17/A11}
\title{\large \bf
Light Higgsino for Gauge Coupling Unification
}
\author{
    Kwang Sik Jeong\footnote{email: ksjeong@pusan.ac.kr}
    }
\affiliation{
Department of Physics, Pusan National University, Busan 46241, Korea
    }

\vspace{2cm}

\begin{abstract}

We explore gauge coupling unification and dark matter in high scale supersymmetry where 
the scale of supersymmetry breaking is much above the weak scale.
The gauge couplings unify as precisely as in low energy supersymmetry if the higgsinos, whose
mass does not break supersymmetry, are much lighter than those obtaining masses from
supersymmetry breaking. 
The dark matter of the universe can then be explained by the neutral higgsino or the gravitino. 
High scale supersymmetry with light higgsinos requires a large Higgs mixing parameter for
electroweak symmetry breaking to take place.
It is thus naturally realized in models where superparticle masses are generated at 
loop level while the Higgs mixing parameter is induced at tree level, like in anomaly 
and gauge mediation of supersymmetry breaking.

\end{abstract}

\maketitle

Supersymmetry (SUSY) is a theoretically well-motivated framework for extending 
the Standard Model (SM), and can be realized in various ways.
Taken as a theoretical guiding principle to physics beyond the SM, naturalness suggests 
low energy SUSY spontaneously broken at a TeV scale or below, for which the minimal 
supersymmetric SM (MSSM) becomes compatible with 
gauge unification~\cite{Dimopoulos:1981yj} 
and provides a viable candidate for cold dark matter~\cite{Jungman:1995df}.
However, the observation of the SM-like Higgs boson with mass near 125 GeV and non-observation 
of new physics signals at the LHC seem to indicate that the scale of SUSY breaking is around 
multi TeV or higher.
One may then need a different approach to the naturalness problem, like for instance
the dynamical relaxation of the weak scale~\cite{Graham:2015cka}.
If realized at a high energy scale, SUSY would play only a partial role in stabilizing the weak scale 
against quantum corrections.\footnote{
In SUSY relaxion models~\cite{  Batell:2015fma,Choi:2015fiu,Evans:2016htp}, 
the higgsino mass parameter $\mu$ receives two or more contributions
whose relative phase is determined by the relaxion, generating the weak scale via cosmological
evolution.  
This mechanism would work also in high scale SUSY with light higgsinos because $|B\mu|^2$,
on which the determinant of the Higgs mass matrix depends, still varies along the relaxion direction.  
Here $B$ denotes the Higgs mixing parameter.
}
Nonetheless it still remains attractive, in particular from the viewpoint of grand unification and 
dark matter.

In this letter we examine if SUSY broken much above the weak scale can account for 
the dark matter of the universe, and retain the unification of gauge couplings in the absence of 
high scale threshold corrections so that extension towards a grand unified theory (GUT) is 
possible.\footnote{See, for instance, Ref.~\cite{Hisano:2013cqa} for threshold corrections 
at the GUT scale in the high scale SUSY scenario.} 
Interestingly, gauge couplings unify as precisely as in low energy SUSY if the
higgsinos, whose mass is not directly connected to SUSY breaking, 
are much lighter than other sparticles and the heavy Higgs doublet~\cite{ArkaniHamed:2005yv,Hall:2011jd}. 
That is, if 
\bea
|\mu| \ll m_{\rm susy},
\eea 
where $\mu$ is the higgsino mass parameter, and $m_{\rm susy}$ denotes the scale of SUSY breaking. 
The lightest sparticle is then the neutral higgsino or the gravitino, which can make up all the dark matter
if produced non-thermally or from thermal scatterings of sparticles. 
 
High scale SUSY with light higgsinos can have interesting phenomenology~\cite{ 
ArkaniHamed:2005yv,Hall:2011jd,Cheung:2005pv,Beylin:2009wz,Fox:2014moa,
Nagata:2014wma,Chun:2016cnm}, 
but needs to address how to achieve electroweak symmetry breaking (EWSB) that requires
a Higgs mixing parameter $B$ much larger than $m_{\rm susy}$. 
We stress that a natural framework for large $B$ is provided by models where 
sparticle masses are generated at loop level while $B$ is induced at tree level, like in
anomaly~\cite{Randall:1998uk,Giudice:1998xp} or gauge 
mediation~\cite{Dine:1994vc,Dine:1995ag}.
We examine the relation between $m_{\rm susy}$ and $\mu$ required for EWSB and gauge unification 
in the scenario where anomaly or gauge mediation is sizable, and estimate the value and scale 
at which the gauge couplings unify.   
  
How light can the higgsinos be compared to other sparticles? 
The most important theoretical constraint on $\mu$ comes from EWSB.  
In the MSSM scalar potential, the squared-mass terms of the neutral Higgs fields read
\bea
&& (m_{H_u}^2+|\mu|^2)|H_u^0|^2 + (m_{H_d}^2+|\mu|^2 )|H_d^0|^2 
\nonumber \\
&&
- (B\mu H_u^0H_d^0 + {\rm h.c.}), 
\eea
including SUSY breaking terms associated with the Higgs doublet fields.
Thus EWSB occurs if
\bea
|B\mu|^2 > (m^2_{H_u} + |\mu|^2) (m^2_{H_d} + |\mu|^2),
\eea 
where  the involved parameters should satisfy 
\bea
2 |B\mu| &<& m^2_{H_u}  + m^2_{H_d} + 2 |\mu|^2,
\eea 
for the scalar potential to be stable along the $D$-flat direction, $|H^0_u|=|H^0_d|$.
It is obvious that the above condition cannot be satisfied if $\mu$ is much larger than $m_{\rm susy}$
as inferred from that it gives a supersymmetric mass to the Higgs bosons.  
For the conventional scenario with  
\bea
\label{EWSB-1}
|\mu| \sim |B| \sim \sqrt{m^2_{H_d}} \sim m_{\rm susy},
\eea 
EWSB is triggered when renormalization group (RG) flow drives $m^2_{H_u}$ to 
negative or small values at low energy, which is possible as it is considerably
affected by loop effects associated with the large top Yukawa coupling. 
 
It is remarkable that EWSB is possible also for $|\mu|\ll m_{\rm susy}$ if the Higgs mixing 
parameter $B$ has a large value around $m^2_{\rm susy}/\mu$.
To be in more detail, the minimization condition
\bea
\sin2\beta &=&
\frac{2|B\mu|}{m^2_{H_u} + m^2_{H_d} + 2|\mu|^2}, 
\nonumber \\
\frac{1}{2}m^2_Z &=&
- |\mu|^2
+ \frac{m^2_{H_d} -m^2_{H_u}\tan^2\beta}{\tan^2\beta -1},
\eea 
shows that EWSB can be achieved when the higgsinos are relatively light and 
the Higgs sector parameters have the hierarchy~\cite{Jeong:2011sg},
\bea
\label{Higgs-hierarchy}
|\mu|  \ll   m_{\rm susy}  \ll |B| \approx \frac{m^2_{\rm susy}}{|\mu|\tan\beta},
\eea
for moderate and large $\tan\beta$, while $m^2_{H_u}$ is negative or much smaller than 
$m^2_{\rm susy}$ in size as in the conventional scenario. 
Here $\tan\beta = \langle |H^0_u| \rangle / \langle |H^0_d| \rangle$, and
$m_Z$ is the $Z$-boson mass.

The hierarchy between the Higgs mixing parameter and sparticle masses 
is naturally realized in a class of SUSY breaking models where sparticles obtain masses at loop
level, that is, for instance, models in which anomaly or gauge mediation is sizable.
A large $B$ in such models has been regarded as problematic in low energy SUSY 
because it makes EWSB difficult unless $\mu$ is below 100~GeV, which is in conflict with
the LEP bound on the chargino mass. 
The situation however changes in high scale SUSY. 

In supergravity, anomaly mediation always induces sparticle masses radiatively, but
the Higgs mixing parameter associated with $\mu$ arises at tree-level:
\bea
m^2_{H_d}|_{\rm AM} &\sim& \frac{m^2_{3/2}}{(8\pi^2)^2},
\nonumber \\
B|_{\rm AM} &\sim& m_{3/2},
\eea
where $m_{3/2}$ is the gravitino mass.
Hence $m_{\rm susy}\ll |B|$ is obtained in models where anomaly mediation is sizable 
to other mediations in size.
In such a case, the gravitino is quite heavy as would be required to make the gravitino
decay before big bang nucleosynthesis (BBN).  
Note also that the higgsino is likely the lightest supersymmetric particle (LSP) because
EWSB requires $|\mu|\ll m_{\rm susy}$.

Another possible way to get a large Higgs mixing parameter is to consider models with
sizable gauge mediation where $\mu$ is dynamically generated from the superpotential term
\bea
\label{GM-mu}
f(X) H_u H_d,  
\eea
for some function $f$.
Here $X$ is the SUSY breaking field that provides masses to messengers.
One may instead consider $f(X,X^*)H_uH_d$ in the K$\ddot{\rm a}$hler potential.
The sparticles then obtain masses at loop level, while the Higgs mixing parameter is
generated at tree level:
\bea
m^2_{H_d}|_{\rm GM} &\sim& \frac{(F/M)^2}{(8\pi^2)^2},
\nonumber \\
B|_{\rm GM} &\sim& F/M,
\eea
for $\langle X \rangle=M+\theta^2 F$.
On the other hand, the vanishing cosmological constant puts a lower bound
on the gravitino mass
\bea
m_{3/2} \gtrsim \frac{M}{M_{Pl}} F/M,
\eea
with $M_{Pl}$ being the reduced Planck mass.
The gravitino can be the LSP if the messenger scale $M$ is intermediate or low.
 
To summarize, there naturally arises a hierarchy between the Higgs mixing parameter 
and the sparticle masses:
\bea
|B| \lesssim 8\pi^2 m_{\rm susy}.
\eea 
in models where anomaly or gauge mediation gives sizable contributions to sparticle 
masses.
For large $B$, we need $|\mu|\ll m_{\rm susy}$ to trigger EWSB.  
In this class of models, therefore, the lightest ordinary supersymmetric particle 
would be a nearly pure higgsino, while the gravitino can be very heavy or light. 

Let us now turn to the issue of gauge coupling unification. 
The MSSM leads to quantitative unification of gauge couplings for low scale SUSY where 
the sparticles are around TeV.
The unified gauge coupling and unification scale are approximately given by 
\bea
\label{gut-TeV}
\frac{1}{g^2_{\rm GUT}|_{\rm TeV} } &\simeq& 2,
\nonumber \\
M_{\rm GUT}|_{\rm TeV} &\simeq& 2\times 10^{16}\, {\rm GeV}.
\eea
Let us consider a simple case in which all the sparticles and the heavy Higgs doublet 
have a common mass.
Then, there exists a value of the common sparticle mass 
\bea
m_\ast
\eea
at which the three gauge couplings meet exactly at one point within the MSSM.
The value of $m_\ast$ is estimated to be  
\bea
m_\ast \approx 1\,{\rm TeV},
\eea
where the dependence on $\tan\beta$, which arises at the two-loop level, is very mild.

To see how gauge coupling unification is affected by sparticle masses, we use one-loop RG evolution 
of gauge couplings in the dimensional reduction scheme under the assumption, for simplicity, 
that the sleptons (squarks) have a universal mass $m_{\tilde \ell}$ ($m_{\tilde q}$). 
From the fact that gauge unification works in low scale SUSY, the unified gauge coupling
is found to be 
\bea
\frac{1}{g^2_{\rm GUT}} &=& \frac{1}{g^2_{\rm GUT}|_{\rm TeV}}
+ \frac{b_a}{8\pi^2} \ln\left(\frac{M_{\rm GUT}|_{\rm TeV}}{M_{\rm GUT}}\right)
\nonumber \\
&&
+\, \sum_\phi \frac{b^\phi_a}{8\pi^2}\ln\left(\frac{m_\phi}{m_\ast}\right),
\eea
for $a=1,2,3$ and summing over the sparticles and heavy Higgs doublet.
Here $b_a$ denotes the $\beta$ function coefficient in the MSSM, and $b^\phi_a$ is
the contribution of the particle $\phi$ to it.
Using the above relations, one finds that successful unification is maintained 
for a sparticle spectrum satisfying
\bea 
\label{unification-condition}
\frac{|\mu|}{m_\ast}
\left(\frac{m_H}{m_\ast}\right)^{ \frac{1}{4}}
\left(\frac{M_2}{m_\ast}\right)^{\frac{1}{3}}
\left(\frac{M_2}{M_3}\right)^{\frac{7}{3}} 
\left(\frac{m_{\tilde \ell}}{m_{\tilde q}}\right)^{\frac{1}{4}}
= 1,
\eea
with $m_\ast \approx 1$~TeV~\cite{Langacker:1992rq,Carena:1993ag,Roszkowski:1995cn,Raby:2009sf,Krippendorf:2013dqa}. 
Here $M_2$ and $M_3$ are the wino and gluino mass, respectively,
and $m_H$ is the heavy Higgs doublet mass.
The unification condition is sensitive to the masses of the higgsino, wino and gluino.  
But it is relatively insensitive to the masses of the heavy Higgs doublet and sfermions.  
In particular, gauge unification becomes independent of the sfermion spectrum if the squarks and 
sleptons have a common mass, which reflects the fact that they form complete 
SU$(5)$ multiplets.
This implies that unification works even when the sfermions are very heavy while other 
sparticles are around $m_\ast$ as in the split SUSY~\cite{ArkaniHamed:2004fb}.
It is also worth noting that the inclusion of two-loop corrections leads to small shift of the scale
$m_\ast$, and does not change the qualitative features.

Under the condition Eq.~(\ref{unification-condition}), the three gauge 
couplings converge to a common value  
\bea
\label{gut-value-scale1}
&& 
\frac{1}{g^2_{\rm GUT} } =  \frac{1}{g^2_{\rm GUT}|_{\rm TeV} }   
\nonumber  \\
&&
\quad
+\,
\frac{1}{8\pi^2} 
\ln \left[
\left( \frac{|\mu|}{m_H} \right)^{  \frac{10}{19} }
\left( \frac{M_2}{|\mu|}  \right)^{ \frac{12}{19} }
\left( \frac {m_{\tilde \ell}} {|\mu|}\right)^{2 }
\right]
\nonumber \\
&& 
\quad 
+\,
\frac{1}{8\pi^2}  \ln \left[
\left( \frac{M_3}{M_2} \right)^{ \frac{125}{19} } 
\left( \frac{m_{\tilde q}}{m_{\tilde \ell}} \right)^{\frac{173}{76} }
\right],
\eea
at a GUT scale
\bea
\label{gut-value-scale2}
M_{\rm GUT} &=&
M_{\rm GUT}|_{\rm TeV}  
\left(\frac{m_H}{|\mu|}\right)^{ \frac{2}{57} } 
\left(\frac{|\mu|}{M_2}\right)^{ \frac{10}{57} }
\nonumber \\
&&
\hspace{0.8cm}
\times
\left(\frac{M_2}{M_3}\right)^{ \frac{25}{57} }
\left(\frac{m_{\tilde \ell}}{m_{\tilde q}}\right)^{ \frac{9}{76} }. 
\eea  
For higgsinos much lighter than other sparticles, the GUT scale and the value of 
unified gauge coupling will get smaller.
If the MSSM is embedded in a GUT, operators mediated by heavy gauge bosons 
around $M_{\rm GUT}$ would induce proton decay mainly via 
$p \to \pi^0 e^+$~\cite{Hisano:2000dg} .
The experimental limits on proton lifetime are evaded for
\bea
\left( \frac{g^2_{\rm GUT}}{0.5} \right)^{-1} 
\left( \frac{M_{\rm GUT}}{2\times 10^{16}{\rm GeV}} \right)^2
\gtrsim 0.1,
\eea 
which is combined with the relations, Eqs.~(\ref{gut-value-scale1}) and (\ref{gut-value-scale2}),
to put an upper bound on the ratio between  $m_{\rm susy}$ and $\mu$:
\bea
\frac{m_{\rm susy}} {|\mu|}  \lesssim 0.6\times 10^4
\left(\frac{M_3}{M_2}\right)^{-\frac{25}{8}},
\eea
for a simplified case where the wino, sfermions and heavy Higgs doublet have 
a similar mass about $m_{\rm susy}$.
The constraint is satisfied in most of the parameter region of our interest. 
The gauge coupling at $M_{\rm GUT}$ increases if there are gauge messengers, enhancing
the proton decay rate~\cite{Hisano:2012wq,Hisano:2015ala}.
As a consequence, the upper bound on $m_{\rm susy}/|\mu|$ presented above is reduced, 
for instance, by a factor of about 0.6 (0.2) for $N_{\rm mess}=1$ (3) and $M_{\rm mess}=10^8$~GeV.
Here $N_{\rm mess}$ is the number of gauge messengers in ${\bold 5}+\bar{\bold 5}$ of SU$(5)$, 
and $M_{\rm mess}$ is their mass.
Note that there are also model-dependent dimension-five operators leading to proton decay
mainly via $p\to K^+ \bar\nu$, which are induced by the exchange of colored Higgs 
multiplets~\cite{Murayama:2001ur}.
The lifetime is approximately proportional to 
$\frac{M^2_{\rm GUT}}{\tan^2\beta} \frac{m^4_{\rm susy}}{M^2_2+\mu^2}$ 
for sfermions around $m_{\rm susy}$, and it follows that high scale SUSY with $m_{\rm susy}$ above 10~TeV
and low $\tan\beta$ is favored to evade the experimental bound~\cite{Hisano:2013exa}.

For gauginos, sfermions and the heavy Higgs doublet around $m_{\rm susy}$, 
the unification condition Eq.~(\ref{unification-condition}) is reduced roughly to
\bea
m_{\rm susy}
\approx
\left(\frac{m_\ast}{|\mu|}\right)^{\frac{12}{7}}
\left(\frac{M_3}{M_2}\right)^{4}
m_\ast,
\eea
where we have kept the dependence on the ratio between the wino and gluino mass as it can
be important. 
The above shows that the sparticle masses $m_{\rm susy}$ can be larger 
than $m_\ast$, i.e.~above TeV, if the higgsinos are much lighter than other sparticles.
The value of $m_{\rm susy}$ consistent with unification is further pushed up if the gluino 
is heavier than the wino~\cite{Wells:2004di}. 

Let us examine if gauge unification works for the Higgs sector with the hierarchy 
of Eq.~(\ref{Higgs-hierarchy}).
Combined with gauge unification, the low energy gaugino masses have a distinctive
and robust pattern~\cite{Choi:2007ka} 
in many mediation schemes including anomaly, gauge, gravity mediation and mixed mediations,
\bea
\left.\frac{M_a}{g^2_a}\right|_{m_{\rm susy}}
\propto\, 1 + b_a \alpha,
\eea
at the one-loop level, whereas scalars have a quite model-dependent mass pattern.
Here $\alpha$ represents the relative importance of anomaly-mediated contributions,
and it is positive in many mixed mediation 
models~\cite{Pomarol:1999ie,Rattazzi:1999qg,Choi:2011xt}.  
Negative $\alpha$ is also possible but in rather involved models~\cite{Okada:2002mv}. 
From the gaugino mass pattern, it follows
\bea
\frac{M_3}{M_2} = 
\frac{1+ b_3 \alpha}{1+b_2 \alpha} \frac{g^2_3}{g^2_2},
\eea
where one would need 
\bea
|\alpha| \leq {\cal O}(1),
\eea
because pure anomaly mediation suffers from the tachyonic slepton problem.
The gauge coupling ratio $g^2_3/g^2_2$ is equal to 3 around TeV, and it decreases
to 1 as energy scale increases. 
For other soft SUSY breaking parameters, one obtains
\bea
\frac{|B|}{m^2_{H_d}} = \frac{\kappa}{m_{\rm susy}},
\eea
for $\kappa$ lying in the range
\bea
0 < \kappa \lesssim 8\pi^2.
\eea 
The parameter $\kappa$ is much larger than order unity if  anomaly and/or gauge mediation
provide sizable masses to sparticles while generating $B$ at tree level. 
Finally, from the minimization condition, the value of $\mu$ appropriate for EWSB 
reads 
\bea
\label{EWSB-mu}
|\mu| \approx \frac{m_{\rm susy}}{\kappa  \tan\beta},
\eea 
and it should not exceed much $m_{\rm susy}$.
As noticed above, the value of $m_{\rm susy}$ consistent with gauge unification 
becomes higher than TeV for $|\mu|\ll m_{\rm susy}$, and is further pushed up 
if the gluino is heavy relative to the wino.  

\begin{figure}[t]
\begin{center}
\begin{tabular}{l}
\includegraphics[trim=0.4cm 0.5cm -0.4cm 0, width=0.5\textwidth]{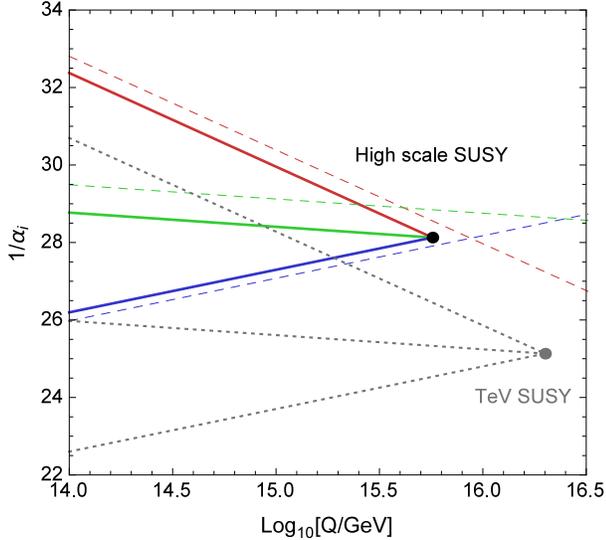}
\end{tabular}
\end{center}
\caption{ 
RG flow of gauge couplings $\alpha_i=g^2_i/4\pi$ in the MSSM.
Dotted gray lines correspond to the running of SU$(3)_C$, SU$(2)_L$ and U$(1)_Y$, respectively,
for low energy SUSY where the heavy Higgs doublet and all sparticles are degenerate around TeV.  
Colored lines show how gauge coupling unification is affected by sparticle masses in high scale
SUSY where the heavy Higgs doublet and sfermions are degenerate at $m_{\rm susy}=200$~TeV.
Colored solid lines are obtained for $M_2=m_{\rm susy}$, $M_3=2m_{\rm susy}$ and $\mu=230$~GeV,
while dashed ones are for $M_2=M_3=\mu=m_{\rm susy}$. 
}
\label{fig:unification}
\end{figure}

We now perform a simple numerical discussion of the sparticle spectrum 
required for EWSB and gauge unification. 
The EWSB relation Eq.~(\ref{EWSB-mu}) is combined with the unification condition 
Eq.~(\ref{unification-condition}) to uniquely fix the values of $\mu$ and $m_{\rm susy}$:  
\bea 
\label{msusy-mu}
\frac{m_{\rm susy}}{m_\ast} &\approx& 
0.1 \times 10^3
\left(\frac{\mbox{\small$M_3/M_2$}}{2}\right)^{\frac{28}{19}}
\left(\frac{\kappa \tan\beta}{300}\right)^{\frac{12}{19}},
\nonumber \\
\frac{|\mu|}{m_\ast} &\approx& 
0.4
\left(\frac{\mbox{\small$M_3/M_2$}}{2}\right)^{\frac{28}{19}}
\left(\frac{\kappa \tan\beta}{300}\right)^{-\frac{7}{12}},
\eea 
for $\kappa\tan\beta \gtrsim 1$, and $\alpha$ of order unity.
Here we have taken into account that low $\tan\beta$ is favored to accommodate 
the 125-GeV Higgs boson in high scale SUSY, for instance, $\tan\beta$ smaller than $4$ 
for $m_{\rm susy}$ above a few tens TeV~\cite{Bagnaschi:2014rsa}, 
and that one has $g^2_3/g^2_2= 2$ around 100~TeV. 
We emphasize that high scale SUSY can be reconciled with gauge unification 
when the higgsinos are much lighter than other sparticles.  
For instance, EWSB and unification are achieved for
$\mu$ around a few hundred GeV and $m_{\rm susy}=10\,$--$\,100$~TeV.
If $M_3/M_2$ gets larger, the required value of $\mu$ and $m_{\rm susy}$
increase by the same factor. 

\begin{figure}[t]
\begin{center}
\begin{tabular}{l}
\includegraphics[trim=0.4cm 0.5cm -0.4cm 0, width=0.5\textwidth]{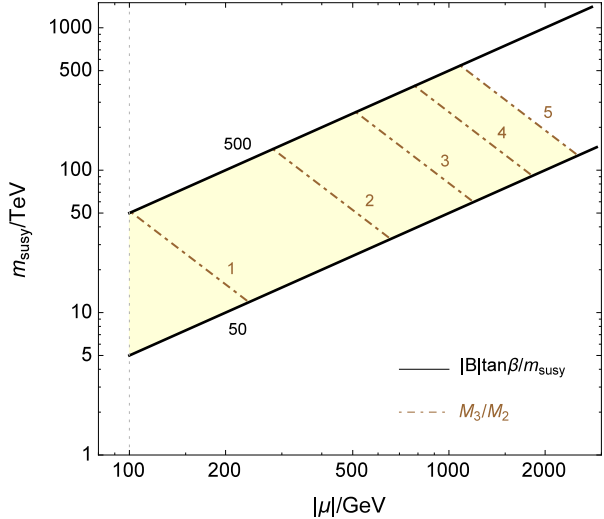}
\end{tabular}
\end{center}
\caption{ 
High scale SUSY with light higgsinos for gauge coupling unification.  
The heavy Higgs doublet and sparticles except for higgsinos have masses
around $m_{\rm susy}$.
For $m_{\rm susy}\gg |\mu|$, EWSB requires a large Higgs mixing parameter,
$|B|\tan\beta \approx m^2_{\rm susy}/|\mu|$.
Here we have taken $50\leq |B|\tan\beta/m_{\rm susy} \leq 500$ for models
where anomaly or gauge mediation is sizable, taking into account 
that high scale SUSY above a few tens TeV can accommodate the observed 125-GeV 
Higgs boson for $\tan\beta\lesssim 4$.
Then, EWSB occurs in the region between the two thick lines, and light higgsinos 
lead to successful gauge coupling unification in the shaded region for $0.5\leq M_3/M_2 \leq 5$. 
}
\label{fig:region}
\end{figure}

Fig.~\ref{fig:unification} shows RG flow of gauge couplings in the MSSM.
High scale SUSY with light higgsinos can lead to gauge coupling unification, where
the three gauge coupling unify as precisely as in the conventional TeV SUSY.  
In Fig.~\ref{fig:region}, the shaded region is compatible with gauge unification.
Here we have used the fact that the SUSY particle mass $m_{\rm susy}$ and
the higgsino mass parameter $\mu$ are fixed according to Eq.~(\ref{msusy-mu}) 
once the EWSB and unification conditions are imposed.  

Finally we discuss dark matter and collider signs in high scale SUSY under consideration.
The LSP, which is stable under $R$-parity conservation, is the higgsino or the gravitino.
For $|\mu|\ll m_{\rm susy}$, the lightest neutralino and chargino are mostly pure higgsino,
and are nearly mass degenerate:  
\bea
\Delta m \equiv m_{\chi^+_1} - m_{\chi^0_1} 
= \Delta m_{\rm tree} + \Delta m_{\rm loop},
\eea
where the tree-level contribution is due to mixing with the bino and wino, and
is positive unless the bino and wino mass have a different sign
\bea
|\Delta m_{\rm tree}|  \simeq 
30{\rm MeV} \left( \frac{10^5  {\rm GeV}}{M_2} \right) 
\left| 1 + 0.3\frac{M_2}{M_1} \right|,
\eea
while the radiative mass difference comes mainly from gauge boson loops~\cite{Thomas:1998wy},
and is approximated by 
\bea
\Delta m_{\rm loop}  \approx  260{\rm MeV} 
\left(\frac{|\mu|}{100{\rm GeV}}\right)^{0.15},
\eea
for $\mu$ below about TeV.
Hence, the mass difference is expected to be positive and larger than the pion mass, 
for which the lightest chargino dominantly decays to the lightest neutralino and the charged pion.
The decay time of the lightest chargino is about 
$0.3\times 10^{-10} {\rm sec} \times (\Delta m/300{\rm MeV})^{-3}$ for $\Delta m$ 
not close to the pion mass.
It would thus be difficult to probe at the LHC, but $e^+e^- \to \gamma \chi^0_1 \chi^0_2$ 
or $\gamma \chi^+_1 \chi^-_1$ mediated by virtual $Z$ boson may provide a visible signal
in future linear colliders.

We first examine the gravitino LSP case, $m_{3/2} < |\mu|$.
In this case, the gravitino production from thermal scatterings can generate the right dark matter
density.
If the freeze-out temperature of the gravitino~\cite{Fujii:2002fv}
\bea
T_f \approx {10^{11}}{\rm GeV} 
\left( \frac{m_{3/2}}{10{\rm GeV}} \right)^2
\left( \frac{100{\rm TeV}}{M_3} \right)^2
\eea
is higher than the reheating temperature $T_{\rm reh}$ after inflation,
the gravitino relic abundance is determined by   
\bea
\Omega_{3/2} h^2 
&\simeq&
0.3 \left(\frac{T_{\rm reh} }{10^9 {\rm GeV} } \right)
\nonumber \\
&&
\times 
\left( \frac{10 {\rm GeV} }{m_{3/2}} \right) 
\left( \frac{M_3} {100 {\rm TeV} }\right)^2,
\eea 
where the gravitino should be heavier than about $100$~keV to be a cold dark 
matter~\cite{Bolz:2000fu}.
The reheating temperature producing the observed dark matter density can be higher than 
about $10^9$~GeV as required for standard thermal leptogenesis~\cite{Fukugita:1986hr}.  
One should however note that the next to lightest supersymmetric particle (NLSP), which
is the neutral higgsino, decays into the gravitino and ordinary particles with a width
\bea
\Gamma_{\chi^0_1} \sim \frac{1}{16\pi} \frac{|\mu|^5}{m^2_{3/2} M^2_{Pl}},
\eea
and so it occurs during or after the BBN epoch in the parameter region of our interest 
while producing high energetic electromagnetic and hadronic showers.
Such late-time decay of the NLSP can significantly alter the abundances of light elements,
spoiling the success of BBN.
A simple way to avoid this cosmological difficulty is to consider $R$-parity violation.
One possibility is to add $R$-parity violating terms that violate the lepton number as 
well but preserve the baryon number to forbid dangerous proton decay 
operators~\cite{Takayama:2000uz,Buchmuller:2007ui}.
For instance, $\Delta W = \mu_i L_i H_u$ with $\frac{\mu_i}{\mu}\tan\beta$ larger than about $10^{-12}$ 
can allow the NLSP to decay very shortly into ordinary particles
while making the gravitino live long enough to make up the dark matter of the universe.
Note that one needs $\frac{\mu_i}{\mu}\tan\beta  \lesssim 10^{-5}$ to avoid washout of baryon 
asymmetry before the electroweak phase transition~\cite{Buchmuller:2007ui,Endo:2009cv}.

Let us move on to the case of $m_{3/2} > |\mu|$ with $R$-parity conservation, 
for which the LSP is the neutral higgsino.
The LSP thermal relic density is significant only for $\mu$ above 1~TeV. 
For smaller $\mu$, the right dark matter density can be generated via non-thermal LSP 
production.
Here we consider a scenario where gravitinos are abundantly produced from the decay
of a heavy scalar field such as inflaton, Polonyi field, or string moduli.
The gravitino decay width reads
\bea
\Gamma_{3/2} = \frac{193}{384\pi} \frac{m^3_{3/2}}{M^2_{Pl}},
\eea
for $m_{3/2}\gg m_{\rm susy}$, and it corresponds to the decay temperature  
\bea
T_{3/2} \simeq 0.25{\rm GeV}
\left(\frac{10}{g_\ast(T_{3/2})}\right)^{\frac{1}{4}} 
\left(\frac{m_{3/2}}{{\rm PeV}}\right)^{\frac{3}{2}},
\eea
where $g_\ast$ counts the effective number of relativistic degrees of freedom.
Thus, LSPs are produced below the LSP freeze-out temperature $\sim |\mu|/20$.
If the gravitino abundance is large enough to make annihilation among produced
LSPs effective, the LSP relic density becomes independent of the initial gravitino
abundance~\cite{Jeong:2011sg,Olive:1990qm}:
\bea
\Omega_{\chi^0_1} h^2& \simeq & 
\frac{0.06}{0.84 c_Z + c_W  }
\left(\frac{g_\ast(T_{3/2})}{10}\right)^{\frac{1}{4}} 
\nonumber \\
&&
\times 
\left(\frac{|\mu|}{100{\rm GeV}}\right)^3
\left(\frac{{\rm PeV}}{m_{3/2}}\right)^{\frac{3}{2}},
\eea 
where $c_Z \equiv (1-m^2_Z/|\mu|^2)^{3/2}/(2-m^2_Z/|\mu|^2)^2$, 
and $c_W$ is similarly defined by taking $m_Z\to m_W$. 
Note that the gravitino should be heavier than about 40~TeV to decay before nucleosynthesis, 
and its mass is bounded from above, $m_{3/2} \lesssim 8\pi^2 m_{\rm susy}$, because 
anomaly mediation is a model-independent source of sparticle masses.
The direct detection of the higgsino dark matter would be challenging since it has only small couplings to
the SM Higgs boson and $Z$ boson suppressed by $m_{W}/M_{1,2}$, making
both spin-dependent and independent scatterings too small to be detected by current experiments. 
However, indirect detection via the process $\chi^0_1 \chi^0_1 \to \gamma\gamma$ or $\gamma Z$
may be possible in future experiments.

\begin{figure}[t]
\begin{center}
\begin{tabular}{l}
\includegraphics[trim=0.4cm 0.5cm -0.4cm 0, width=0.5\textwidth]{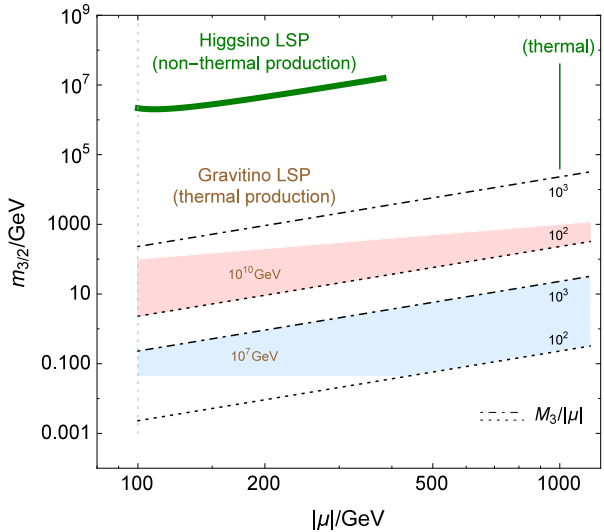}
\end{tabular}
\end{center}
\caption{ 
Dark matter in high scale SUSY with light higgsinos.
For the higgsino LSP case with $50 \leq m_{\rm susy}/|\mu| \leq 500$, 
non-thermal LSP production from heavy gravitino decays yields the right dark matter density
along the thick green line while satisfying the condition $m_{3/2} < 8\pi^2 m_{\rm susy}$.  
The correct density can be obtained also from the higgsino thermal relic for $\mu$ about 1~TeV,
which is along the thin vertical green line, where the gravitino mass is bounded by 
 $m_{3/2} < 8\pi^2 m_{\rm susy}$ and the BBN constraint.  
On the other hand, in the case of a gravitino LSP, gravitinos produced by thermal scatterings can explain
the observed dark matter density in the light red and blue shaded region for a reheating temperature
indicated by the numbers.  
Here we have taken $10^2 \leq M_3/|\mu| \leq 10^3$ and the reheating temperature 
$T_{\rm reh}=10^{10}$~GeV ($10^7$~GeV) for the region between the upper (lower) dot-dashed and
dotted lines.
The non-shaded region between the upper (lower) dot-dashed and dotted lines leads 
to $m_{3/2}>|\mu|$ ($T_{\rm reh}>T_f$), where $T_f$ is the freeze-out temperature of
the gravitino.  
}
\label{fig:DM}
\end{figure}

Fig.~\ref{fig:DM} shows the region where the right dark matter density
$\Omega_{\rm DM} h^2 =0.12$ is obtained.
If the neutral higgsino is the LSP, one can consider non-thermal production of higgsinos from 
heavy gravitino decays.
On the other hand, in the case of a gravitino LSP, gravitino production by thermal scatterings
can yield the right relic density.  
 
We have studied gauge coupling unification and dark matter in high scale SUSY where 
sparticles have masses much higher than the weak scale as would 
be indicated by tensions between low energy SUSY and the LHC results.
High scale SUSY admits gauge coupling unification as exactly as in low energy SUSY
if the higgsinos are much lighter than those obtaining SUSY breaking masses.
In this scenario,  the LSP is the neutral higgsino or the gravitino, which can make up all the dark 
matter of the universe if produced non-thermally in the former case and thermally from sparticle
scatterings in the latter case. 
A large Higgs mixing parameter required for EWSB is naturally obtained in models where 
sparticle masses are generated at loop level while the Higgs mixing parameter arises at tree
level, as is the case in anomaly and gauge mediation.
Then, gauge coupling unification and dark matter would indicate high scale SUSY around
10~TeV -- a PeV and light higgsinos below a few TeV. 

\section*{Acknowledgment}
The author thanks to Bumseok Kyae and Fuminobu Takahashi for useful discussions. 
This work is supported by the National Research Foundation of Korea (NRF) grant funded by the Korea government (MSIP) (NRF-2015R1D1A3A01019746).

\end{document}